\documentclass[aps, prd, superscriptaddress, nofootinbib, preprintnumbers, twocolumn]{revtex4}
\usepackage{graphics}
\usepackage{epsfig,amsfonts,amssymb,amsmath}
\usepackage{epstopdf}
\usepackage{latexsym}
\usepackage{colordvi}
\usepackage{refstyle}
\usepackage{float}
\newcommand{\be}{\begin{equation}}
\newcommand{\ee}{\end{equation}}
\newcommand{\bea}{\begin{eqnarray}}
\newcommand{\eea}{\end{eqnarray}}
\begin{document}

\title{Holographic conductivity of 1+1 dimensional systems in soft wall model}
\author{Neha Bhatnagar}
\email{bhtngr.neha@gmail.com}
\author{Sanjay Siwach}%
 \email{sksiwach@hotmail.com}
\affiliation{%
 Department of Physics,Banaras Hindu University,\\Varanasi-221005, India\\
}%

\begin{abstract}
We study the optical conductivity of 1+1 dimensional systems using soft wall model in the bottom up approach of  AdS/CFT (anti-de Sitter/conformal field theory) duality. We find the numerical results for optical conductivity and investigate the system using holographic model in the probe limit. The dependence of conductivity on chemical potential is also investigated. Further, we extend the soft wall model as a `no-wall' model by eliminating the dilaton background and study the response of the system in a simplified approach.\\

{\bf Keywords:}  AdS/CFT duality; soft wall model; no-wall model
\end{abstract}

 \pacs{04.70.s ; 11.25.Tq}

\maketitle

\section{\label{sec:level1}Introduction}

The failure of perturbative method and the limitations of traditional non-perturbative approaches makes it difficult to study strongly correlated systems. Lattice field theory\cite{2489469}, is an effective method to study the static equilibrium at high temperature and low density, but  requires very high performance computational techniques. However, with the introduction of AdS/CFT correspondence by Maldacena\cite{Maldacena:1997re} the situation turns  favorable. The correspondence relates $\mathcal{N}=4$ super Yang-Mills (SYM) conformal field theory in four-dimensions with the Type {\it II B} string theory on $AdS_5 \times S^5$ spacetime. The generalisation of this conjecture known as gauge/gravity duality or holography and is used to obtain many significant results about the thermodynamical and hydrodynamical aspects of strongly coupled systems using their weakly coupled gravity dual\cite{Witten:1998zw,Son:2002sd,Policastro:2002se,Son:2007vk,Ge:2008ak,Iqbal:2008by,Jain:2009bi,Bechi:2009aq,Faulkner:2010jy,Mandal:2011uq,Li:2014txa}.  
Using the classical gravity solutions in the bulk, we can study holographic theory for the gauged system on the boundary. These observations  further leads to the development in understanding of various phenomena like phase transition, Hall effect, Nerst effect and other experimentally observed properties in a strongly correlated condensed matter systems\cite{Gubser:2005ih,Gubser:2008px,Hartnoll:2007ai,Hartnoll:2008vx,Konoplya:2009hv,Nishioka:2009zj,Basu:2009vv,Horowitz:2010jq,Pan:2010at,Davison:2013jba,Blake:2013bqa,Donos:2014uba,Zeng:2014uoa,Cai:2015jta,Cai:2015wfa}. 

In this work, we study the optical conductivity of a 1+1 dimensional system using its gravity dual BTZ black hole in 2+1 dimensions\cite{Banados:1992wn} in the soft wall model\cite{Erlich:2005qh,Karch:2006pv,Erlich:2009me}. This system has been investigated recently\cite{Hung:2009qk,Maity:2009zz,Ren:2010ha,Balasubramanian:2010sc,Nurmagambetov:2011yt,Liu:2011fy,Bu:2012qr,Hendi:2016pvx} and the transport properties have been studied both numerically and analytically using charged scalar field coupled with the gauge field in the gravity action. The objective of our work is to calculate the transport properties of the system in much simpler form using  soft-wall model\cite{Afonin:2015fga,Afonin:2010hn}. It has been shown in \cite{Afonin:2015fga} that different dilaton profile corresponds to different condensates for the system. Although, in this present work we have not investigated the phase transition explicitly, but the pattern for the optical conductivity indicates the metallic phase for 1+1 dimensional system. Using the soft wall model, first we investigate the optical conductivity for the chargeless black hole. Then, the real and imaginary part of the optical conductivity are obtained for different value of the chemical potentials in the `no wall' model by eliminating the dilaton background in front of the Ricci curvature in the action \cite{Afonin:2015fga,Bhatnagar:2015uma}.

\section*{Holographic Setup}
Let us consider the soft wall model in 2+1 dimensional Einstein-Maxwell system given as,
\begin{equation}
S=\int d^{3}x\sqrt{-g} e^{-2\phi}\left(\frac{1}{2\kappa^2}(R-2\Lambda)+\frac{1}{4g^2}F^2\right)
\end{equation}
where $F^2 $ is field strength of U(1) gauge field $A_\mu$,$\kappa^2 =8\pi G_3=1$ ($G_3$ is the three dimensional Newton's constant) and $\Lambda$(cosmological constant)$=-2/L^2$ (taking L=1 which is a AdS length scale).\\
The equations of motion using the above action are given as,
\begin{eqnarray}
R_{\mu\nu}-\frac{1}{2}g_{\mu\nu}R-\Lambda g_{\mu\nu}&=&8\pi G_3 T_{\mu\nu}\\
\nabla_\mu e^{-2\phi}F^{\mu\nu}&=&0 
\end{eqnarray}
with $T_{\mu\nu}$ the energy-momentum tensor. The solution of the equations of motion is a charged black hole system and is given by the following metric ansatz,
\begin{equation}
ds^2=\frac{1}{z^2 }\left(-f(z)dt^2+dx^2+\frac{dz^2}{f(z)}\right)
\end{equation}
where
\begin{eqnarray} 
A_t=\mu\ln\frac{z}{z_h},~~~~
f(z)=1-\frac{z^2}{z_h^2}+\frac{1}{2} \mu ^2 z^2 \ln\frac{z}{z_h}
\end{eqnarray}
and $\mu$ is the chemical potential.
Further we set $z\rightarrow 0$ as the boundary of $AdS_3$ and $z=z_h$ is taken to be the horizon of the black hole with a limit $z_h \rightarrow 1$ used in this paper.
The Hawking temperature of the black hole is given by, 
\begin{equation}
T=\frac{f'(z)}{4\pi}\vline_{z \rightarrow z_h} =\frac{4-\mu^2}{8\pi}
\end{equation}
The gauge field perturbation used for the calculation is given as 
\begin{equation}
 \textit{A}_{x}(z,t)=\tilde{A}_{x}(z) e^{-i\omega t}
\end{equation}
Then, the optical conductivity is obtained using the Ohm's law,
\begin{equation}\label{eq:cond}
\sigma(\omega)~=~\frac{J^x}{E}~=~\frac{J^x}{i\omega A_{x}^0}
\end{equation}
where $A_x^0$ act as the source and $J^x$ is the response for the boundary theory.
After scaling the metric perturbatins as $\tilde{h}_{mn}=e^{2\phi}{h}_{mn}$ we get following equations of motion (at $\mu=0$).
\begin{equation}
\label{eq:gauge}
\tilde{A}_x''+ \left(\frac{f'}{f}-2\phi'\right)\tilde{A}_x'+\frac{\omega ^2}{f^2}\tilde{A}_x= 0
\end{equation}

Here we have studied two different models for the dilaton profile, Model I $(\phi=z)$ and Model-II $(\phi=z^2)$ for studying the 1+1 dimensional system in the probe limit. Since it is believed that the dilaton profile is emerging from the condensation of the scalar field, the two different shapes are analogous to the type-II and type-I coherence factors of the holographic superconductor \cite{Afonin:2015fga}.

The numerical results of the optical conductivity using equation (\ref{eq:gauge}) has been shown in Fig.1. where the behavior of the optical conductivity matches with the literature \cite{Hartnoll:2008vx,Afonin:2015fga}.  
\begin{figure}[h!]
\begin{center}
\includegraphics[width=0.35\textwidth]{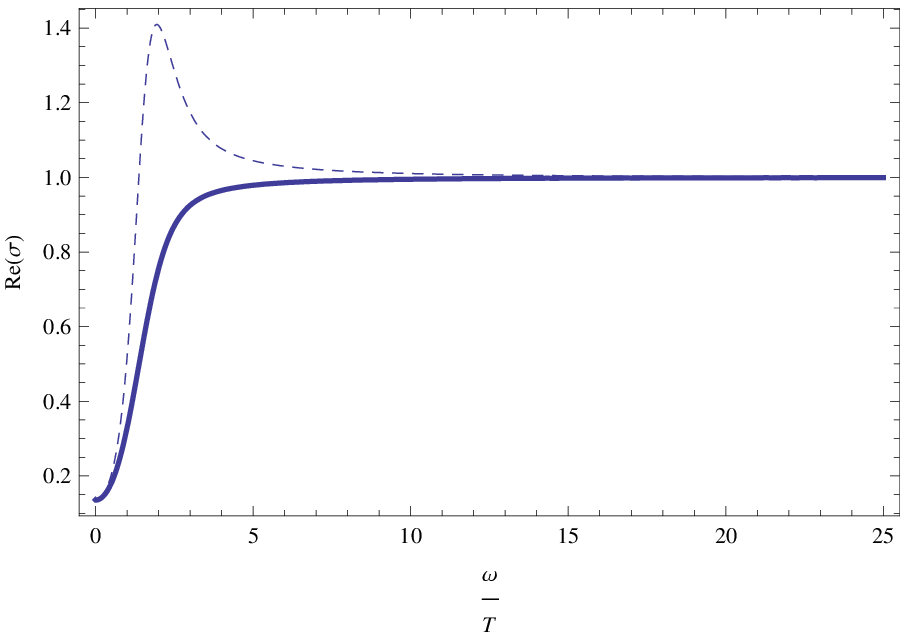}
\includegraphics[width=0.35\textwidth]{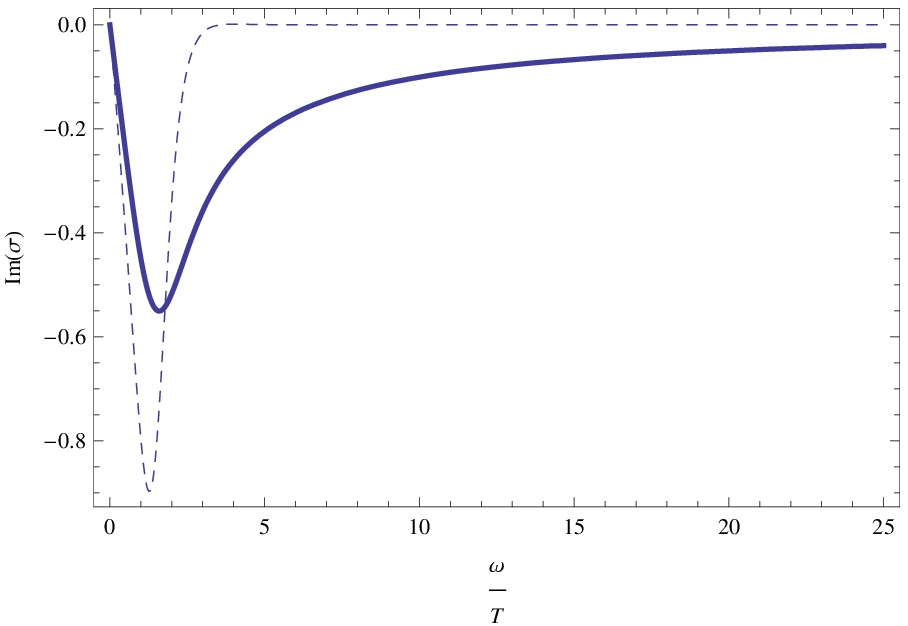}
\end{center}
\caption{Frequency response of optical conductivity at $\mu=0$ for Model-I ($\phi=z$(thick)) and Model-II ($\phi=z^2 $(dashed)).}
\end{figure}

\section*{Optical conductivity for charged black hole in 2+1 dimensions}
We study the optical conductivity with chemical potential in the soft wall model. For convenience, in addition to the scaling of metric perturbation, the gauge field has been redefined as, $\tilde{A_x}=e^{\phi}A_x$. This scaling will eliminate the dilaton background and the model works like a `no-wall' model. Introducing the metric and gauge field perturbations as,
\begin{eqnarray}
& g_{mn}(z,t)= h_{mn}(z)e^{-i\omega t}\\
& \textit{A}_{m}(z,t)=\tilde{A}_{m}(z) e^{-i\omega t}
\end{eqnarray}
We obtain the equation of motion  (in the gauge $A_r=0$ and eliminating $h_{tx}$ using its equations of motion),  
\begin{eqnarray}
A_x''+ \left(\frac{f'}{f}+\frac{1}{z}\right)A_x'+\nonumber\\ \left(\phi''-\phi'^2+\phi' \left(\frac{f'}{f}+\frac{1}{z}\right)-\frac{{A'_t}^2 z^2 e^{\phi}}{f}+\frac{\omega ^2}{f^2}\right)A_x= 0
\end{eqnarray}

In  the probe limit of the charged black hole solution, the temporal component of the gauge field at the horizon is given by 
$A_t=\mu \ln[\frac{z}{z_h}]$. Using the ingoing boundary condition, 
\begin{equation}
A_x\vline_{z=1}~(1-z^2)^{-i\omega/2}+...
\end{equation}
Optical conductivity is given by \cite{Maity:2009zz,Ren:2010ha,Balasubramanian:2010sc},
\begin{equation}
\sigma(\omega)=\frac{-(A_x-zA'_x \ln[z])}{z A'_x}\vline_{z\rightarrow \epsilon}
\end{equation}
Introducing a small value of chemical potential in the probe limit (taking $f(z)=1-z^2$) we study the frequency response of the  conductivity in Fig. 2. The presence of Drude peak has been observed for both the models at low frequency indictaing the metallic phase of the system. 

\begin{figure}[h!]
\includegraphics[width=0.37\textwidth]{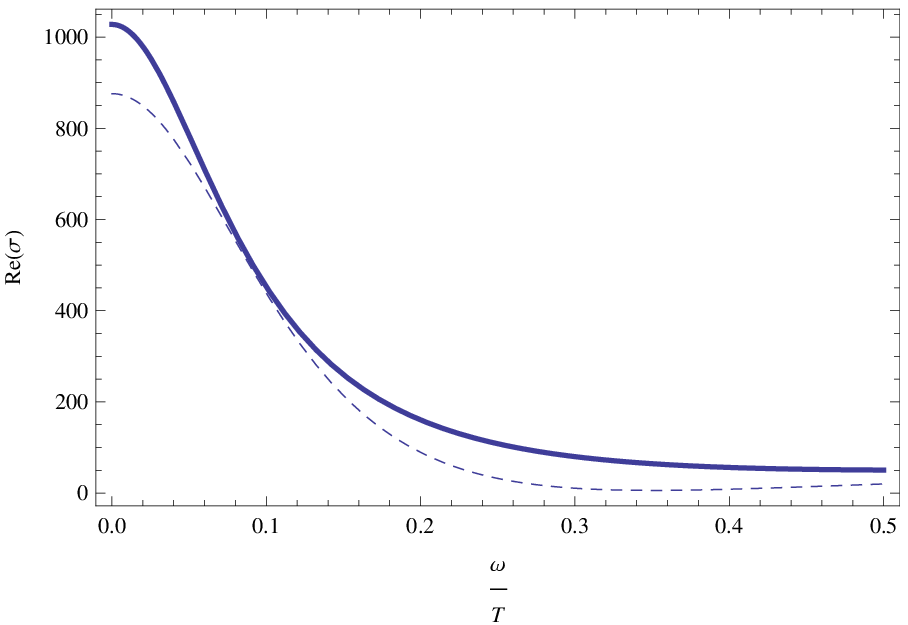}
\includegraphics[width=0.37\textwidth]{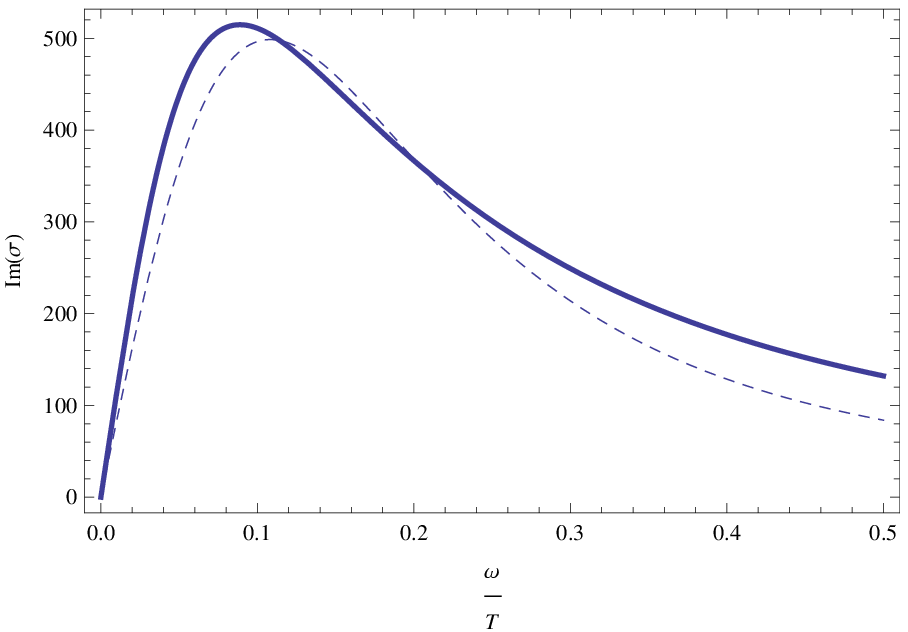}
\caption{Frequency response of conductivity for the metallic phase of 1+1 dimensional system at $\mu=$0.2, for Model-I ($\phi=z$)(thick) and Model-II($\phi=z^2$)(dashed).}
\end{figure}
Further the dependence of conductivity on chemical potential in the probe limit  can be verified from Fig.3 and Fig.4. We see the clear shift in the peak along with the suppression of the conductivity as we increase the chemical potential of the system. We also notice that the flow is same for the different chemical potential at higher frequency.  
\begin{figure}[h!]
\includegraphics[width=0.37\textwidth]{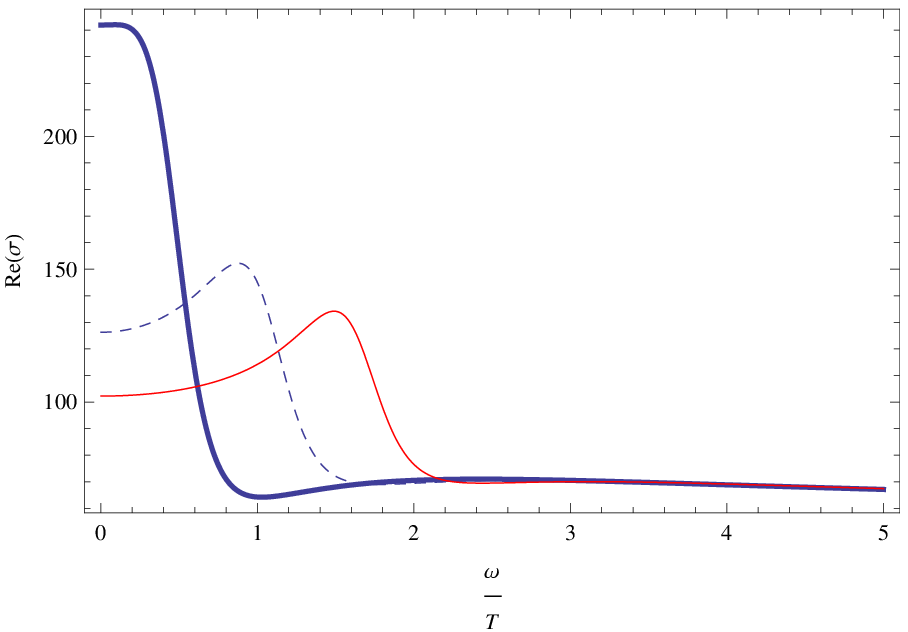}
\includegraphics[width=0.37\textwidth]{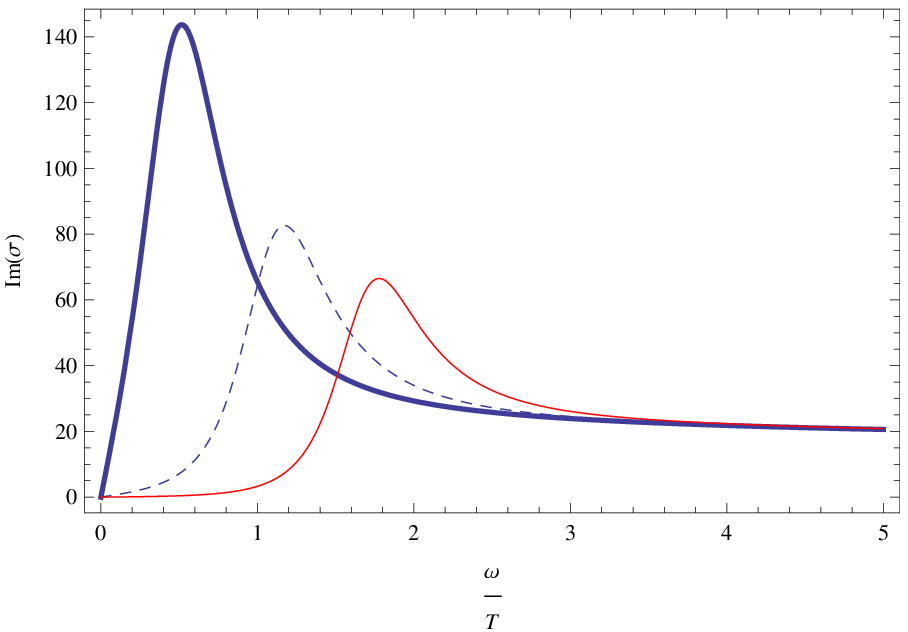}
\caption{Frequency response of conductivity at $\mu=$0.5(thick), 1(dashed), 1.5(red) for Model-I ($\phi=z$).}
\end{figure}

\begin{figure}[h!]
\includegraphics[width=0.37\textwidth]{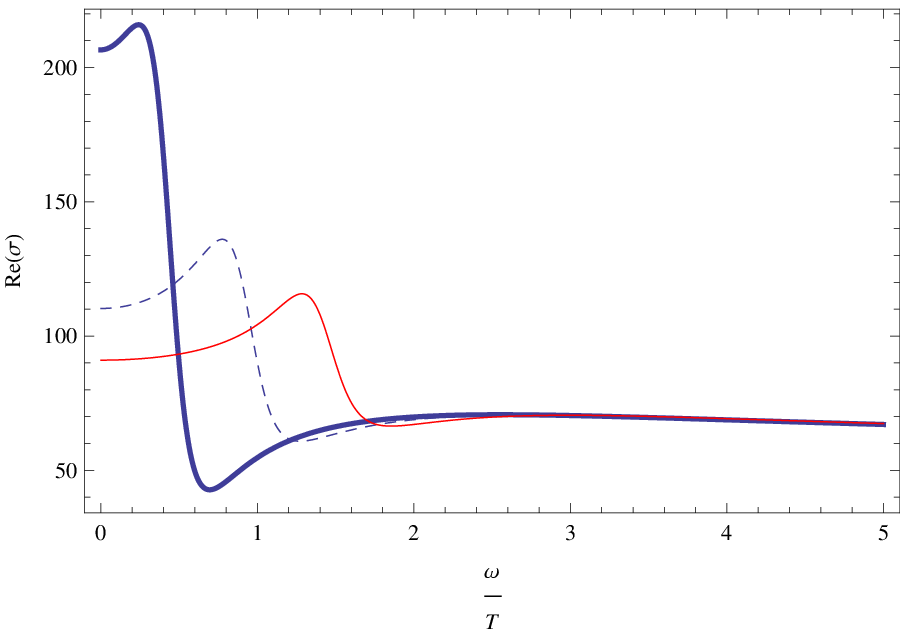}
\includegraphics[width=0.37\textwidth]{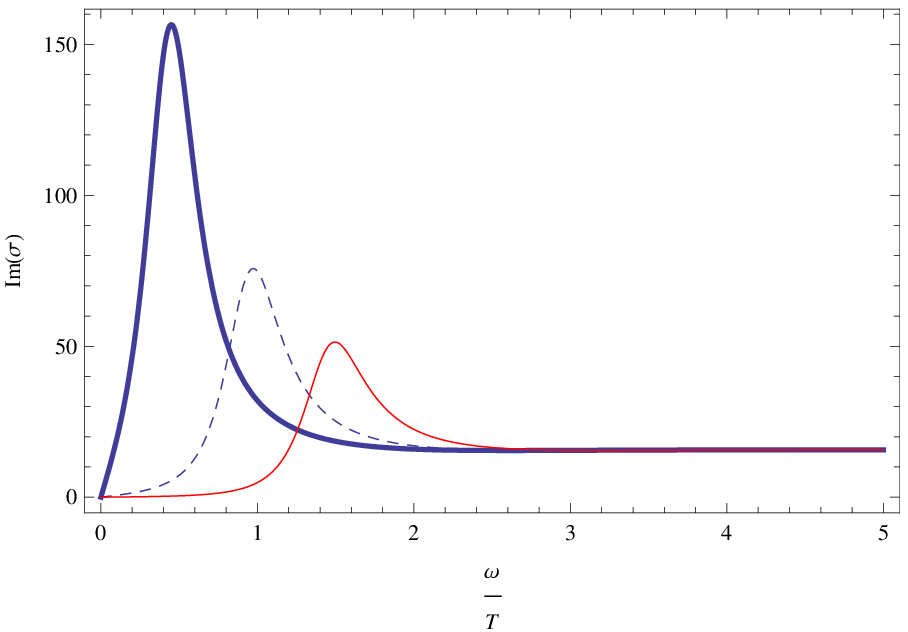}
\caption{Frequency response of optical conductivity at $\mu=$0.5(thick), 1(dashed), 1.5(red) for Model-II ($\phi=z^2 $).}
\end{figure}

Finally, we consider the charged black hole solution with $f(z)=1-\frac{z^2}{z_h^2}+\frac{1}{2} \mu ^2 z^2 \ln\frac{z}{z_h}$ and study the behavior for the real and imaginary part of the optical conductivity at high frequency. It has been noticed earlier that oscillatory behavior indicates the interference effects due to the charge fractionalization \cite{Berg}. The results has been shown in Fig.5 for both the models at fixed chemical potential.

\begin{figure}[htb!]
\includegraphics[width=0.37\textwidth]{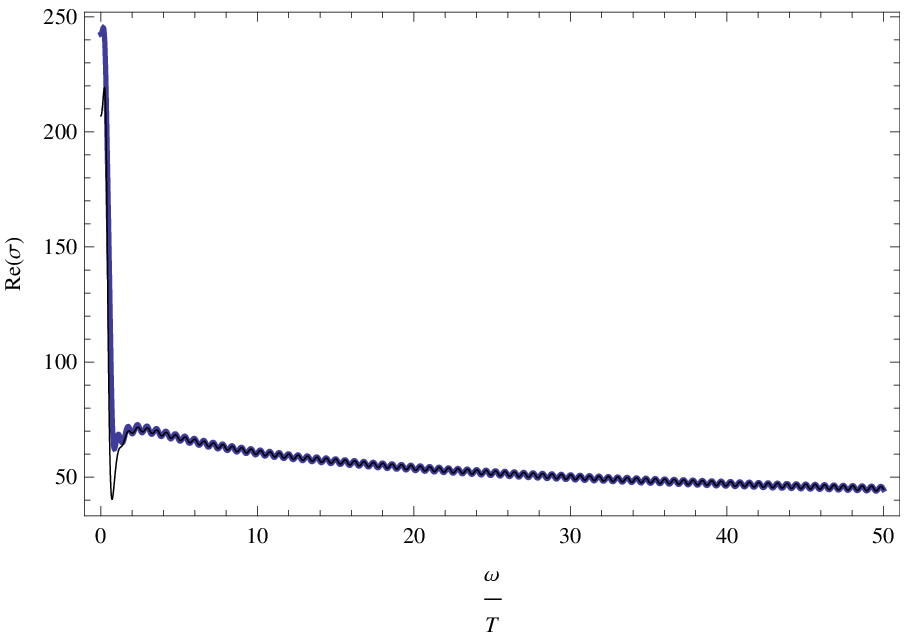}
\includegraphics[width=0.37\textwidth]{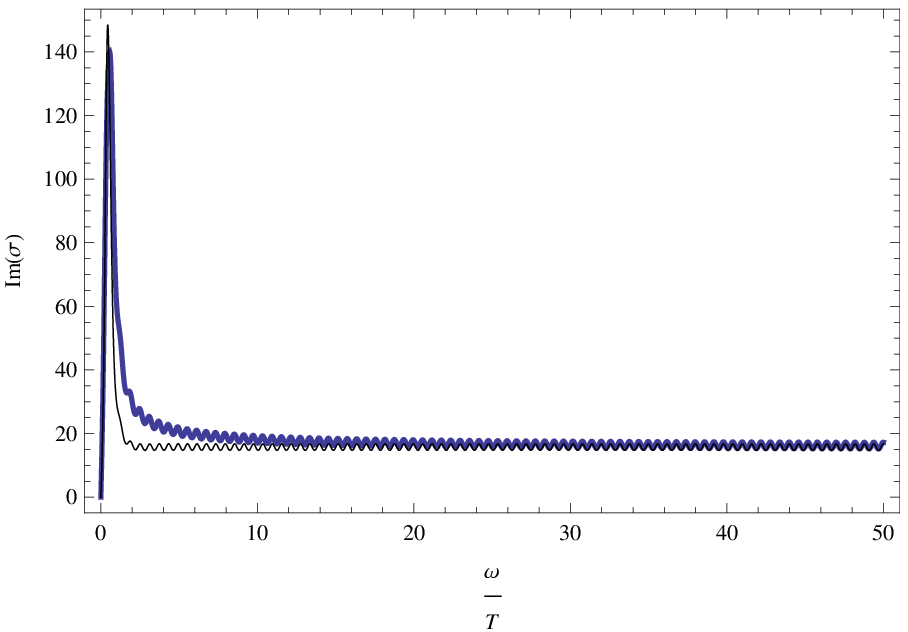}
\caption{Frequency response of conductivity at $\mu=$0.5, for Model-I(blue)($\phi=z$) and Model-II(black) ($\phi=z^2$ )}.
\end{figure}

\section*{Conclusions and Discussions}
We have studied the numerical results of optical conductivity for a 1+1 dimensional system in soft wall model using the charged 2+1 dimensioanl BTZ black hole in the bulk. We have studied two different dilaton profiles to solve the gauge field equation and study the behavior of optical conductivity indicating the  Fermi-Luttinger liquid type behavior\cite{Maity:2009zz}. We also notice the suppression of conductivity peak as we increase the value of chemical potential of the system. It would be interesting to study the phase structure of the above system with chemical potential in order to decide the existence of superconductivity like phase as indicated in the literature\cite{Ren:2010ha}.

\section*{Conflict of Interests}
The authors declares that there is no conflict of interest regarding the publication of this paper.

\section*{Acknowledgements}
We would like to acknowledge Prof. Debanand Sa and Dr. Matteo Baggioli for valuable suggestions and discussions.


\begin{thebibliography}{00}
\bibitem{2489469}
John B.~ Kogut and Mikhail A.~ Stephanov,
"The Phases of Quantum Chromodynamics: From Confinement to Extreme Environments",
Cambridge University Press, 2004.

\bibitem{Maldacena:1997re}
  J.~M.~Maldacena,
``The Large N limit of superconformal field theories and supergravity,''
 Adv.\ Theor.\ Math.\ Phys.\  {\bf 2}, 231 (1998)
 [hep-th/9711200]

\bibitem{Witten:1998zw} 
 E.~Witten,
``Anti-de Sitter space, thermal phase transition, and confinement in gauge theories,''
 Adv.\ Theor.\ Math.\ Phys.\  {\bf 2}, 505 (1998)
[hep-th/9803131].

\bibitem{Son:2002sd}
  D.~T.~Son and A.~O.~Starinets,
``Minkowski space correlators in AdS / CFT correspondence: Recipe and applications,''  
JHEP {\bf 0209}, 042 (2002) 
 [hep-th/0205051].

 \bibitem{Policastro:2002se}
 G.~Policastro, D.~T.~Son and A.~O.~Starinets,
'`From AdS / CFT correspondence to hydrodynamics,'' 
 JHEP {\bf 0209}, 043 (2002) 
 [hep-th/0205052].

\bibitem{Son:2007vk} 
  D.~T.~Son and A.~O.~Starinets,
  ``Viscosity, Black Holes, and Quantum Field Theory,''
  Ann.\ Rev.\ Nucl.\ Part.\ Sci.\ {\bf 57},95 (2007)
  [hep-th/0704.0240].

\bibitem{Ge:2008ak}
 X.~H.~Ge, Y.~Matsuo, F.~W.~Shu, S.~J.~Sin and T.~Tsukioka,
``Density Dependence of Transport Coefficients from Holographic Hydrodynamics,''
Prog.\ Theor.\ Phys.\  {\bf 120}, 833 (2008)
 [hep-th/0806.4460 ].

\bibitem{Iqbal:2008by}
 N.~Iqbal and H.~Liu,
``Universality of the hydrodynamic limit in AdS/CFT and the membrane paradigm,''
 Phys.\ Rev.\ D {\bf 79}, 025023 (2009)
[hep-th/0809.3808].

\bibitem{Jain:2009bi}
 S.~Jain,
'`Universal properties of thermal and electrical conductivity of gauge theory plasmas from holography,''
 JHEP {\bf 1006}, 023 (2010)
[hep-th/0912.2719].

\bibitem{Bechi:2009aq} 
 J.~Bechi,
 ``Comments on the Chiral Symmetry Breaking in Soft Wall Holographic QCD,''
 [hep-th/0912.2681].

\bibitem{Faulkner:2010jy}
 T.~Faulkner, H.~Liu and M.~Rangamani,
``Integrating out geometry: Holographic Wilsonian RG and the membrane paradigm,''
 JHEP {\bf 1108}, 051 (2011)
[hep-th/1010.4036].

\bibitem{Mandal:2011uq} 
  G.~Mandal and T.~Morita,
  ``What is the gravity dual of the confinement/deconfinement transition in holographic QCD?,''
  J.\ Phys.\ Conf.\ Ser.\  {\bf 343}, 012079 (2012)
  [hep-th/1111.5190].

\bibitem{Li:2014txa} 
  D.~Li and M.~Huang,
  ``Dynamical holographic QCD model,''
  EPJ Web Conf.\  {\bf 80}, 00011 (2014)
  [hep-th/1409.8432].


\bibitem{Gubser:2005ih} 
  S.~S.~Gubser,
  ``Phase transitions near black hole horizons,''
  Class.\ Quant.\ Grav.\  {\bf 22}, 5121 (2005)
  [hep-th/0505189].

\bibitem{Gubser:2008px} 
  S.~S.~Gubser,
  ``Breaking an Abelian gauge symmetry near a black hole horizon,''
  Phys.\ Rev.\ D {\bf 78}, 065034 (2008)
 [hep-th/0801.2977].

\bibitem{Hartnoll:2007ai} 
  S.~A.~Hartnoll and P.~Kovtun,
  ``Hall conductivity from dyonic black holes,''
  Phys.\ Rev.\ D {\bf 76}, 066001 (2007)
  [hep-th/0704.1160].

\bibitem{Hartnoll:2008vx} 
  S.~A.~Hartnoll, C.~P.~Herzog and G.~T.~Horowitz,
  ``Building a Holographic Superconductor,''
  Phys.\ Rev.\ Lett.\  {\bf 101}, 031601 (2008)
  [hep-th/0803.3295].

\bibitem{Konoplya:2009hv} 
  R.~A.~Konoplya and A.~Zhidenko,
  ``Holographic conductivity of zero temperature superconductors,''
  Phys.\ Lett.\ B {\bf 686}, 199 (2010)
  [hep-th/0909.2138].

\bibitem{Nishioka:2009zj} 
  T.~Nishioka, S.~Ryu and T.~Takayanagi,
  ``Holographic Superconductor/Insulator Transition at Zero Temperature,''
  JHEP {\bf 1003}, 131 (2010)
  [hep-th/0911.0962].

\bibitem{Basu:2009vv} 
  P.~Basu, J.~He, A.~Mukherjee and H.~H.~Shieh,
  ``Hard-gapped Holographic Superconductors,''
  Phys.\ Lett.\ B {\bf 689}, 45 (2010)
  [hep-th/0911.4999].

\bibitem{Horowitz:2010jq} 
  G.~T.~Horowitz and B.~Way,
  ``Complete Phase Diagrams for a Holographic Superconductor/Insulator System,''
  JHEP {\bf 1011}, 011 (2010)
  [hep-th/1007.3714].

\bibitem{Pan:2010at} 
 Q.~Pan and B.~Wang,
 ``General holographic superconductor models with Gauss-Bonnet corrections,''
 Phys.\ Lett.\ B {\bf 693}, 159 (2010)
 [hep-th/1005.4743].

\bibitem{Davison:2013jba} 
  R.~A.~Davison,
  ``Momentum relaxation in holographic massive gravity,''
  Phys.\ Rev.\ D {\bf 88}, 086003 (2013)
  [hep-th/1306.5792].

\bibitem{Blake:2013bqa} 
  M.~Blake and D.~Tong,
  ``Universal Resistivity from Holographic Massive Gravity,''
  Phys.\ Rev.\ D {\bf 88}, no. 10, 106004 (2013)
  [hep-th/1308.4970].

\bibitem{Donos:2014uba} 
  A.~Donos and J.~P.~Gauntlett,
  ``Novel metals and insulators from holography,''
  JHEP {\bf 1406}, 007 (2014)
 [hep-th/1401.5077].

\bibitem{Zeng:2014uoa} 
  H.~B.~Zeng and J.~P.~Wu,
  ``Holographic superconductors from the massive gravity,''
  Phys.\ Rev.\ D {\bf 90}, no. 4, 046001 (2014)
  [hep-th/1404.5321].

\bibitem{Cai:2015jta} 
  R.~G.~Cai, R.~Q.~Yang, Y.~B.~Wu and C.~Y.~Zhang,
  ``Massive $2$-form field and holographic ferromagnetic phase transition,''
  JHEP {\bf 1511}, 021 (2015)
  [hep-th/1507.00546].

\bibitem{Cai:2015wfa} 
  R.~G.~Cai and R.~Q.~Yang,
  ``Insulator/metal phase transition and colossal magnetoresistance in holographic model,''
  [hep-th/1507.03105].

\bibitem{Banados:1992wn} 
  M.~Banados, C.~Teitelboim and J.~Zanelli,
  ``The Black hole in three-dimensional space-time,''
  Phys.\ Rev.\ Lett.\  {\bf 69}, 1849 (1992)
  [hep-th/9204099].

\bibitem{Erlich:2005qh}
  J.~Erlich, E.~Katz, D.~T.~Son and M.~A.~Stephanov,
``QCD and a holographic model of hadrons,''
  Phys.\ Rev.\ Lett.\  {\bf 95}, 261602 (2005)
 [hep-ph/0501128].

\bibitem{Karch:2006pv}
 A.~Karch, E.~Katz, D.~T.~Son and M.~A.~Stephanov,
``Linear confinement and AdS/QCD,''
 Phys.\ Rev.\ D {\bf 74}, 015005 (2006)
 [hep-ph/0602229].

\bibitem{Erlich:2009me}
 J.~Erlich,
 ``How Well Does AdS/QCD Describe QCD?,''
 Int.\ J.\ Mod.\ Phys.\ A {\bf 25}, 411 (2010)
 [hep-th/0908.0312].


\bibitem{Hung:2009qk} 
  L.~Y.~Hung and A.~Sinha,
  ``Holographic quantum liquids in 1+1 dimensions,''
  JHEP {\bf 1001}, 114 (2010)
  [hep-th/0909.3526].

\bibitem{Maity:2009zz} 
  D.~Maity, S.~Sarkar, N.~Sircar, B.~Sathiapalan and R.~Shankar,
  ``Properties of CFTs dual to Charged BTZ black-hole,''
  Nucl.\ Phys.\ B {\bf 839}, 526 (2010)
 [hep-th/0909.4051].

\bibitem{Ren:2010ha} 
  J.~Ren,
  ``One-dimensional holographic superconductor from AdS$_3$/CFT$_2$ correspondence,''
  JHEP {\bf 1011}, 055 (2010)
  [hep-th/1008.3904].

\bibitem{Balasubramanian:2010sc} 
  V.~Balasubramanian, I.~Garcia-Etxebarria, F.~Larsen and J.~Simon,
  ``Helical Luttinger Liquids and Three Dimensional Black Holes,''
  Phys.\ Rev.\ D {\bf 84}, 126012 (2011)
[hep-th/1012.4363].

\bibitem{Nurmagambetov:2011yt} 
  A.~J.~Nurmagambetov,
  ``Analytical approach to phase transitions in rotating and non-rotating 2D holographic superconductors,''
[hep-th/1107.2909].

\bibitem{Liu:2011fy} 
  Y.~Liu, Q.~Pan and B.~Wang,
  ``Holographic superconductor developed in BTZ black hole background with backreactions,''
  Phys.\ Lett.\ B {\bf 702}, 94 (2011)
 [hep-th/1106.4353].

\bibitem{Bu:2012qr} 
  Y.~Bu,
  ``1+1-dimensional p-wave superconductors from intersecting D-branes,''
  Phys.\ Rev.\ D {\bf 86}, 106005 (2012)
  [hep-th/1205.1614].


\bibitem{Hendi:2016pvx} 
  S.~H.~Hendi, B.~E.~Panah and S.~Panahiyan,
  ``Massive charged BTZ black holes in asymptotically (a)dS spacetimes,''
  JHEP {\bf 1605}, 029 (2016)
  [hep-th/1604.00370].


\bibitem{Afonin:2015fga} 
  S.~S.~Afonin and I.~V.~Pusenkov,
  ``Soft wall model for a holographic superconductor,''
  Eur.\ Phys.\ J.\ C {\bf 76}, no. 6, 342 (2016)
  [hep-th/1506.05381].

\bibitem{Afonin:2010hn} 
  S.~S.~Afonin,
  ``No-Wall Holographic Model for QCD,''
  Int.\ J.\ Mod.\ Phys.\ A {\bf 26}, 3615 (2011)
 [hep-th/1012.5065].


\bibitem{Bhatnagar:2015uma} 
  N.~Bhatnagar and S.~Siwach,
  ``RG flow of AC Conductivity in Soft Wall Model of QCD,''
  Int.\ J.\ Mod.\ Phys.\ A {\bf 31}, no. 08, 1650030 (2016)
  [hep-th/1512.06523].

\bibitem{Berg}
E.Berg, Y.Oreg, E.-A.Kim and F.von Oppen,
"Fractional Charge on an Integer Quantum Hall Edge"',
Phys. Rev. Letter.{\bf102},236402.

\end{thebibliography}
\end{document}